\documentclass[manuscript,onefignum,onetabnum]{siamonline171218}

\usepackage{booktabs} % For formal tables

\usepackage{wrapfig}
\usepackage{amsmath}
\usepackage{bm}
\usepackage{multirow}
\usepackage{tcolorbox}
\usepackage{eurosym}
\tcbuselibrary{theorems}
\newtcbtheorem[number within=section]{mydefinition}{Model}%
{colback=yellow!5,colframe=yellow!35!black,fonttitle=\bfseries}{th}
\newtcbtheorem[number within=section]{mydefinition2}{Definition}%
{colback=blue!5,colframe=blue!35!black,fonttitle=\bfseries}{th}
\newtcbtheorem[number within=section]{mydefinition3}{Rule of Thumb}%
{colback=red!5,colframe=red!35!black,fonttitle=\bfseries}{th}
\usepackage{epsf}
\usepackage{epsfig}
\usepackage{mathrsfs}
\usepackage{subfigure}
\usepackage{graphics}
\usepackage{booktabs}
\usepackage{balance}
\usepackage{graphicx}
\usepackage{bbm}
\usepackage{balance}
\usepackage{epstopdf}
\usepackage{pifont}

\usepackage{amsfonts}

\usepackage{pifont}

\newcommand{\block}{b}
\newcommand{\fatalities}{\phi}
\newcommand{\stayHome}{h}
\newcommand{\population}{\mathcal{P}}

\newcommand{\dd}{\delta}
\usepackage{tikz}
\usepackage{pgfkeys}
\usepackage{xcolor}
\usepackage{epstopdf}
\usepackage{etoolbox}
\usepackage{tcolorbox}
\usepackage{tabularx}
\usepackage{array}
\usepackage{colortbl}
\tcbuselibrary{skins}
\usetikzlibrary{automata}
\usetikzlibrary{shapes,arrows, backgrounds}
\usetikzlibrary{mindmap}
\usetikzlibrary{positioning}
\usetikzlibrary{decorations.pathreplacing}

\definecolor{stateColor}{RGB}{0,128,0}
\definecolor{green}{RGB}{0,128,0} 
\definecolor{UK}{RGB}{0,51,0} 
\definecolor{US}{RGB}{0,204,0} 
\definecolor{vpColor}{RGB}{0,5,0}
\tikzstyle{date} = [rectangle, fill=stateColor!0,text centered, rounded corners, text = green, rotate=0]
\tikzstyle{eraBox} = [rectangle, fill= green!10,text centered,  text = green, rotate=0]

\DeclareRobustCommand{\officialeuro}{%
  \ifmmode\expandafter\text\fi
  {\fontencoding{U}\fontfamily{eurosym}\selectfont e}}

\newcolumntype{Y}{>{\raggedleft\arraybackslash}X}

\tcbset{tab1/.style={fonttitle=\bfseries\large,fontupper=\footnotesize\sffamily,
colback=yellow!10!white,colframe=red!75!black,colbacktitle=maroon!40!white,
coltitle=black,center title,freelance,frame code={
\foreach \n in {north east,north west,south east,south west}
{\path [fill=red!75!black] (interior.\n) circle (3mm); };},}}

\tcbset{tab2/.style={enhanced,fonttitle=\bfseries,fontupper=\footnotesize\sffamily,
colback=yellow!10!white,colframe=red!50!black,colbacktitle=blue!40!white,
coltitle=black,center title}}

\tcbset{tab3/.style={enhanced,fonttitle=\bfseries,fontupper=\footnotesize\sffamily,
colback=yellow!10!white,colframe=red!50!black,colbacktitle=red!40!white,
coltitle=black,center title}}
\usepackage{wrapfig}
\usepackage{times}
\usepackage{lipsum}
\usepackage{amsfonts}
\usepackage{graphicx}
\usepackage{epstopdf}
\usepackage{algorithmic}
\ifpdf
  \DeclareGraphicsExtensions{.eps,.pdf,.png,.jpg}
\else
  \DeclareGraphicsExtensions{.eps}
\fi

\usepackage{enumitem}
\setlist[enumerate]{leftmargin=.5in}
\setlist[itemize]{leftmargin=.5in}

\newsiamremark{remark}{Remark}
\newsiamremark{hypothesis}{Hypothesis}
\crefname{hypothesis}{Hypothesis}{Hypotheses}
\newsiamthm{claim}{Claim}

\headers{Effectiveness and Compliance to Social Distancing During COVID-19}{K. Bushman et al.}

\title{Effectiveness and Compliance to Social Distancing During COVID-19}

\author{Kristi Bushman, Konstantinos Pelechrinis, Alexandros Labrinidis\thanks{School of Computing and Information, University of Pittsburgh 
  (\email{k.bushman@pitt.edu}, \email{kpele@pitt.edu}, \email{labrinid@pitt.edu})}}

\usepackage{amsopn}

\makeatletter
\newcommand*{\addFileDependency}[1]{% argument=file name and extension
  \typeout{(#1)}% latexmk will find this if $recorder=0 (however, in that case, it will ignore #1 if it is a .aux or .pdf file etc and it exists! if it doesn't exist, it will appear in the list of dependents regardless)
  \@addtofilelist{#1}% if you want it to appear in \listfiles, not really necessary and latexmk doesn't use this
  \IfFileExists{#1}{}{\typeout{No file #1.}}% latexmk will find this message if #1 doesn't exist (yet)
}
\makeatother

\newcommand*{\myexternaldocument}[1]{%
    \externaldocument{#1}%
    \addFileDependency{#1.tex}%
    \addFileDependency{#1.aux}%
}
\usepackage{multirow}
\usepackage{helvet}
\usepackage{courier}
\usepackage{epsf}
\usepackage{epsfig}
\usepackage{mathrsfs}
\usepackage{subfigure}
\usepackage{graphics}
\usepackage{booktabs}
\usepackage{balance}
\usepackage{graphicx}
\usepackage{balance}
\usepackage{epstopdf}
\usepackage{pifont}
\usepackage{amsfonts}
\usepackage{pifont}
\usepackage{epstopdf}
\DeclareMathSizes{10}{9}{6}{4}
\usepackage{etoolbox}
\ifpdf
\hypersetup{
  pdftitle={Lineup evaluations through in-game win probability models and Bayesian adjustment},
  pdfauthor={Konstantinos Pelechrinis}
}
\fi

\myexternaldocument{ex_supplement}

\begin{document}

\maketitle

\begin{abstract}
In the absence of pharmaceutical interventions to curb the spread of COVID-19, countries relied on a number of nonpharmaceutical interventions to fight the first wave of the pandemic. 
The most prevalent one has been stay-at-home orders, whose the goal is to limit the physical contact between people, which consequently will reduce the number of secondary infections generated. 
In this work, we use a detailed set of mobility data to evaluate the impact that these interventions had on alleviating the spread of the virus in the US as measured through the COVID-19-related deaths. 
To establish this impact, we use the notion of Granger causality between two time-series. 
We show that there is a unidirectional Granger causality, from the average fraction of people staying completely home weekly to the number of COVID-19-related deaths with a lag of 3 weeks. 
We further analyze the mobility patterns at the census block level to identify which parts of the population might encounter difficulties in adhering and complying with social distancing measures. 
This information is important, since it can consequently drive interventions that aim at helping these parts of the population. 
\end{abstract}

% REQUIRED
%\begin{abstract}
%  This is an example SIAM \LaTeX\ article. This can be %used as a
%  template for new articles.  Abstracts must be able to stand alone
%  and so cannot contain citations to the paper's references,
%  equations, etc.  An abstract must consist of a single paragraph and
%  be concise. Because of online formatting, abstracts must appear as
%  plain as possible. Any equations should be inline.
%\end{abstract}

\section{Introduction}
\label{sec:intro}

Since the first reported case of COVID-19 in early December 2019 in Wuhan, China, the world has experienced dramatic changes in an effort for societies to deal with the pandemic. 
Given the absence of pharmaceutical interventions (i.e., a medical treatment or a vaccine), governments and health officials have relied on non-pharmaceutical interventions, including {\em shelter-at-home} orders, contact tracing and volume testing. 
The reasoning behind shelter-at-home interventions is to limit the physical contacts between people, which furthermore limits the transmission of the virus. 
Given the absence of a vaccine, this does not mean that the virus will be eradicated but rather, limiting people's mobility will allow the health systems to operate under capacity and be as effective as possible, consequently limiting the number of fatalities. 

Of course, these measures have not been without controversy. 
Hence, it is important to examine whether they are effective in achieving their goal. 
Using a granular mobility dataset for the US obtained from SafeGraphs (details provided in the following sections) and COVID-19-related fatalities we show that average fraction of people staying home weekly {\em Granger-causes} the number of COVID-19-related fatalities with a 3-weeks lag. 
We also examined for and did not find any evidence of bidirectional Granger causality, i.e., feedback effects of people altering their mobility as a response to the change of the number of fatalities (e.g., as a reaction to the news). 
We also provide a short-term prediction model for the number of COVID-19 related fatalities in US one-week out, using only information about population-level mobility behavior and fatalities over the past three weeks. 

Given the effectiveness of these measures it is critical to understand who in the population complies and to what extent. 
Differences in compliance levels are not necessarily by choice. 
For example, many people are essential workers and hence, need to spend time outside of their home. 
Others might not have to physically be at work, but they have to take care of family members living in other households etc. 
Identifying these parts of the population can provide critical information on possible policies/interventions that could further increase compliance, without compromising the needs of people. 
Therefore, in this work we build a framework using a beta regression model to predict the percentage of time spent daily at home within a census block based on demographic characteristics. 
Using these models we can then examine various hypotheses on whether specific demographics of interest are associated with a change in mobility above and beyond of what was expected from the mobility patterns prior to stay-at-home orders. 
We focus on two particular demographics, age and race, and show that show that minorities and older people, while significantly increasing their stay at home, this increase is smaller compared to that white and younger people. 
We further provide some possible mechanisms that lead to this observation and show that income disparities can explain a sizable part of this difference. 
%Using these models we can then estimate the difference-in-differences (DiD) for the response variable for the demographics of interest. 
%We further obtain an estimate of the uncertainty for DiD, by resampling the predicted beta distribution from our models. 
%DiD essentially provides a comparison with a counterfactual response, which assumes that the demographics examined do not have any impact on the time spent home daily, and the two corresponding populations follow parallel trends. 
%We showcase how one can apply this framework by considering the relationship between mobility, age and race in the state of Pennsylvania. 
%Analyzing data from the state of Pennsylvania, our results indicate that minorities and population older than 50 years old might find obstacles in spending as much time as home as their white or younger counterparts.  
%
%We would like to note here, that while difference-in-differences regression is traditionally used for estimating quasi-causal effects from observational data \cite{RePEc:tpr:restat:v:67:y:1985:i:4:p:648-60}, this is not how we are using it here. 
%In this work, we simply borrow the high level idea behind the DiD method to obtain a better understanding of the results as we will discuss later in detail. 
The main contributions of our work can be summarized as follows: 

\begin{itemize}
    \item Provide a Granger-causality analysis on the impact of stay-at-home orders on COVID-19-related fatalities
    \item Design a framework for quantifying adherence to social distancing according to various demographics
    \item Design a dynamic dashboard to visualize both the {\em raw} mobility data as well as, the results from our analysis. 
\end{itemize}

We believe that our work can provide critical information to local officials and policy makers. 
The rest of the paper is organized as follows. 
Section \ref{sec:data} provides a description of the data we used for our analysis, as well as, a brief review on related to our study literature. 
Section \ref{sec:granger} provides our Granger-causality analysis, while Section \ref{sec:beta-did-demo} introduces our framework for identifying the relationship between social distancing compliance and demographics. 
We conclude our work and discuss its limitations and directions for future work in Section \ref{sec:conclusions}.

\section{Data and Related Work}
\label{sec:data}

In this section we describe the dataset we use for our analysis, as well as, relevant to our study literature. 
The code for the analysis presented in the paper can be found on our github repository: \url{https://github.com/kpelechrinis/epiDAMIK20-COVID}.

{\bf SafeGraph data: }
SafeGraph has released a detailed mobility dataset based on the locations of about 18 million mobile phones across the US. 
This information is obtained through various mobile applications that partner with SafeGraph. 
This provides diverse population coverage, while the data are provided in an aggregated manner, with steps taken towards satisfying differential privacy requirements. 
While a detailed description can be found on SafeGraph's COVID-19 data consortium page \cite{safegraph-data}, the main information that we will use is the daily mobility patterns for census block groups (CBG). 
In particular, for each day and each census block group since 01/01/2020 we obtain - among other - the following daily information: 

\begin{itemize}
    \item {\tt completely\_home\_device\_count}: This is the number of devices within the CBG of interest that did not leave their home. 
    \item {\tt distance\_traveled\_from\_home}: This is the median distance traveled during the day from all the devices whose home is within the CBG of interest
    \item {\tt median\_percentage\_time\_home}: This is the median percentage of time spent at home during the day from devices whose home is within the CBG of interest
    \item {\tt destination\_cbgs}: This is the CBGs that were visited during the day from devices whose home is within the CBG of interest. Each destination block is also associated with the number of devices in the SafeGraph dataset that performed this transition.
\end{itemize}

{\bf COVID-19 data:} In order to evaluate any (Granger causal) impact between mobility and COVID-19-related fatalities we need to utilize data related to the number of confirmed cases and deaths. 
While an accurate number for the daily number of infections would be the most appropriate variable for this analysis, it is widely known that the reported numbers are a severe undercount of the actual number of infections. 
On the other hand the number of fatalities is also inaccurate but it is considered a more robust signal for the prevalence of the disease. 
Albeit it is a lagged signal, with an average of 15-20 days delay \cite{lauer2020incubation}. 
We obtain our data from the NY Times github repository \cite{nyt-git-covid}.

{\bf COVID-19 and mobility:} 
Excluding clinical interventions (potential treatments, vaccine, etc.), limiting mobility and inter-personal contacts has been the most central intervention in an effort to contain the pandemic. 
As such, several studies have analysed the changes in human mobility across various regions using granular mobility data (e.g., \cite{gao2020mapping,klein2020assessing,pullano2020population,colorado-covid} with the list being non-exhaustive). 
Aleta {\em et al.} \cite{aleta2020modeling} further utilize these mobility information to drive agent-based simulators in order to understand the impact of contact tracing and testing on a possible second wave of the disease. 
Zhang {\em et al.} \cite{zhang2020changes} have further analyzed contact surveys from the early epidemic stage in China and built transmission models to quantify the impact of social distancing and school closures. 
This line of research is of course still developing as restrictions are lifted, new measures potentially coming in the possibility of a second wave etc.

{\bf Public health non-pharmaceutical evaluation:} 
Of course, similar non-pharmaceutical interventions have been applied in the past as well and there is a volume of research that evaluates their impact. 
For example, Ahmed {\em et al.} \cite{ahmed2018effectiveness} provide a review study on social distancing measures in workplace. 
Their review includes both epidemiological as well as, modeling studies and they concluded that overall workplace social distancing reduced the influenza attack rate approximately 23\%. 
Similarly, Rashid {\em et al.} \cite{rashid2015evidence} reviewed studies that evaluated various measures (school closings, work-from-home etc.) for dealing with the 2009 influenza pandemic. 
They identified that workplace interventions provide moderate reduction in transmissions (20-30\%). 
Other non-pharmaceutical interventions include the banning of mass events. 
While intuitively this seems to be a particularly effective strategy, prior literature has shown that this is true only in combination with other interventions \cite{ishola2011could,markel2007nonpharmaceutical}. 
One of the reasons for this is the contact time at such events is relatively small compared to the time spent in schools, workplaces, or other community locations \cite{ferguson2020report}. 
The literature aforementioned is not exhaustive. 
However, to the best of our knowledge, there is no study that uses the notion of Granger causality for non-pharmaceutical interventions. 
Contrary to the majority of existing studies that rely on large-scale simulation models, or, analyzing a small case (e.g., a restaurant, a specific workplace etc.), we take a macroscopic approach, looking at the aggregate adherence to these interventions and the macroscopic results (e.g., total fatalities). 

\section{Evaluating Social Distancing}
\label{sec:granger}

In this section we will begin by introducing the notion of Granger causality between two time series and then we will see how it applies to our case. 

\subsection{Granger Causality}
\label{sec:granger_method}

Granger causality is a statistical test that aims at identifying whether a time-series $x(t)$ provides useful information in predicting timeseries $y(t)$ \cite{granger1969investigating}. 
It is eminent to understand that Granger causality is what Granger himself described, ``temporally related'' or ``predictive causality'', rather than the traditional notion of causality. 
Simply put, $x(t)$ is said to Granger-cause $y(t)$ if it precedes in time and is able to improve the predictions of $y(t)$ beyond auto-regressive models. 
While this might not be a useful notion for what is needed in areas like clinical treatments, it is particularly useful and has been extensively used in econometrics, public policy etc. (e.g., \cite{bollen2011twitter,arvin2015transportation,hood2008two,buiter1984granger,narayan2012energy} with the list being non-exhaustive). 

Formally, the examination of whether $x(t)$ Granger-causes (G-causes for short) $y(t)$ one needs to build the following two models: 

\begin{eqnarray}
M_0: & y(t) = a_{00} + \sum_{i=1}^{m} a_{0i} y(t-i) + \epsilon_0 \label{eq:M0}\\
M_1: & y(t) = a_{10} + \sum_{i=1}^{m} a_{1i} y(t-i) + \sum_{i=1}^{p} b_i x(t-i)+ \epsilon_1 \label{eq:M1}
\end{eqnarray}

The first model (Eq. \ref{eq:M0}) is essentially a pure auto-regressive model on $y$ up to lag $m$ (called the restricted model), while the second one includes lagged terms from the time-series $x(t)$ to be explored as a potential Granger cause (called the unrestricted model). 
Given this setting the following null hypothesis is examined: via conducting an F-test: 

\begin{eqnarray}
H_0: & b_1 = b_2 = \dots = b_p = 0 \label{eq:granger-null}
\end{eqnarray}

The null here is the hypothesis that no explanatory power is jointly added from the lags of $x$. 
So eventually, we retain all the lagged values of $x$ that are individually statistically significant (t-statistic), but in order to reject $H_0$ that $x$ does not G-cause $y$, all these lags need to add explanatory power (as compared to the restricted model). 
We would like to note here that the time series need to be stationary before performing the Granger test. 
Hence, if the original data are not stationary they should be transformed to eliminate the possibility of autocorrelation (e.g., through differentiation). 

\subsection{Shelter-at-home and COVID-19 fatalities}
\label{sec:granger_analysis}

We are interested in examining whether the mobility of people in the US G-causes the number of fatalities from COVID-19. 
Here, we would like to emphasize that for the latter, we are using the number of COVID-19 deaths $\fatalities$ reported from health authorities as discussed in Section \ref{sec:data}. 
We do not make use of any information related to excess fatalities, or any attempt to estimate the under-reporting factor in fatalities. 
For the G-cause variable, we first obtain the fraction of devices in each census block group $\block$ that stayed exclusively at home daily\footnote{We also examined the median percentage of time spent at home, with similar results. In fact, the median percentage of time spent at home and the fraction of people staying completely at home daily are highly correlated, with a correlation coefficient $\rho = 0.93$.} $\stayHome_{\block}$. 
We then obtain a weighted average value over all the CBGs,  $\stayHome_{US}(t)$, for each day $t$, where the weights are the sample size in each block. 
We further aggregate the data weekly, since there are known inconsistencies and delays in reporting cases and deaths. %  \cite{}. 
Weekly aggregation should remove some of the associated noise with COVID-19 daily reports. 

\begin{figure*}
    \centering
    \includegraphics[scale=0.4]{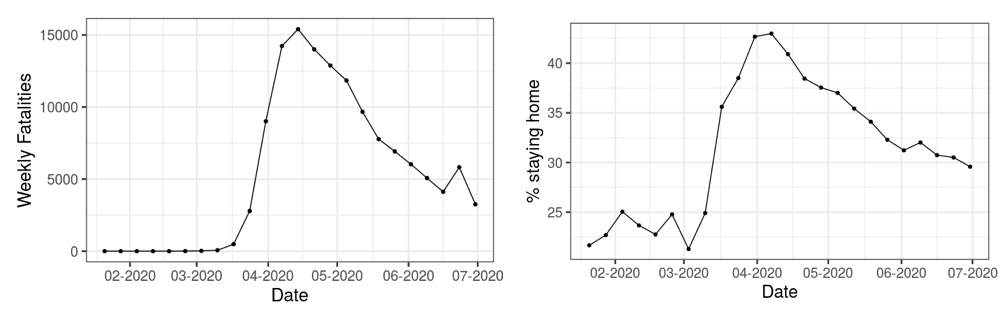}
    \caption{Weekly time series for COVID-19-related fatalities (left) and percentage of people staying at home (right).}
    \label{fig:tseries}
\end{figure*}

Figure \ref{fig:tseries} shows the two weekly time-series of interest for the period between 01/21/2020 (when the COVID-19 cases started being recorded) and 07/03/2020.  
We apply the Kwiatkowski–Phillips– Schmidt–Shin test \cite{kwiatkowski1992testing} and we identify that these time-series are not stationary. 
However, differentiating both time-series will lead to stationarity. 
Running the Granger causality test for lags up to 6 weeks (given the length of our time-series longer lags cannot be tested), we obtain the results in Table \ref{tab:granger_results}. 
As we can see, there is evidence that mobility G-causes COVID-19-related fatalities at a lag of about 2 weeks. 
We also examined for bidirectional G-causality, i.e., people listening to the news and number of fatalities, and reacting with changes in their mobility. 
However, we did not find any supporting evidence.  
%Finally, Table \ref{tab:var_model} presents the auto-regressive model identified from the Granger causality procedure. 

\iffalse
\begin{center}
\begin{table}
\begin{tabular}{ |c|c| } 
\hline
Lag (weeks) & Test p-value  \\
\hline
\hline
 1 & 0.14 \\ 
 2 & 0.05 \\
 3 & 0.11 \\
\hline
\end{tabular}
\vspace{0.1in}
\caption{Test results for mobility Granger-causing COVID-19 fatalities.}
\label{tab:granger_results}
\vspace{-0.2in}
\end{table}
\end{center}
\fi

\iffalse
\begin{table}[htbp]\centering
\begin{tabular}{c c c c c c }
\cmidrule[\heavyrulewidth]{2-6}
 & & & {\bf Lag} & & \\
\cmidrule[\heavyrulewidth]{2-6}
& \textbf{1} & \textbf{2} & \textbf{3} & \textbf{4} & \textbf{5 }\\ 
\midrule
 $b_{1}$         &      $153.3^{*}$  &  $107.4^{*}$ & $106.1^{*}$ & $105.7^{**}$ & $122.6^{*}$\\
 $b_{2}$    &  -   & $174.1^{**}$ & $145.3^{**}$ & $119.2^{**}$ & $181.0^{*}$\\
 $b_{3}$     &   -    & - & -35.1 & -47.8 & 15.1 \\
 $b_{4}$  & - & - & & -65.4 & -101.1 \\
$b_{5}$ & - & - & &-  & -31.9 \\
\midrule
Adjsuted $R^2$ & 0.58  & 0.83 & 0.85 & 0.94 & 0.95 \\
F-test (p-val)          &  3.5 & 6.0   & 6.2 & 7.33  & 5.86   \\
 & (0.07) & (0.008) & (0.007) & (0.003) & (0.003)\\
\bottomrule
\addlinespace[1ex]
\multicolumn{3}{l}{\textsuperscript{***}$p<0.01$, 
  \textsuperscript{**}$p<0.05$, 
  \textsuperscript{*}$p<0.1$}
\end{tabular}
\caption{Individual coefficients' significance and F-test result for various lags.}
\label{tab:granger_results}
\end{table}
\fi

\begin{table}[htbp]\centering
\begin{tabular}{c c c c c c c}
\cmidrule[\heavyrulewidth]{2-7}
 & & & {\bf Lag} & & \\
\cmidrule[\heavyrulewidth]{2-7}
& \textbf{1} & \textbf{2} & \textbf{3} & \textbf{4} & \textbf{5 } & \textbf{6}\\ 
\midrule
 $b_{1}$         &      $263.4^{*}$  &  $156.0$ & $236.9^{*}$ & $230.9^{*}$ & $344.4^{**}$ & $289.5^{*}$\\
 $b_{2}$    &  -   & $401.4^{**}$ & $432.7^{**}$ & $539.6^{**}$ & $447.8^{**}$ & $574.9^{*}$\\
 $b_{3}$     &   -    & - & $348.1^{*}$ & $516.2^{*}$ & $675.9^{**}$ & $760.1^{*}$  \\
 $b_{4}$  & - & - &- & $186.7$ & $-18.1$ & 65.1 \\
$b_{5}$ & - & - & -&-  & 145.6 & $-64.3$ \\
$b_{6}$ &- &- &- &- &- & $-10.3$ \\
\midrule
Adjusted $R^2$ & 0.46  & 0.69 & 0.77 & 0.79 & 0.91 & 0.9 \\
F-test (p-val)          &  5.08 & 10.3   & 11.1 & 18.9  & 8.5&  12.6  \\
 & (0.03) & ($<$0.01) & ($<$0.01) & ($<$0.01) & ($<$0.01)& ($<$0.01) \\
\bottomrule
\addlinespace[1ex]
\multicolumn{3}{l}{\textsuperscript{**}$p<0.01$, 
  \textsuperscript{*}$p<0.05$, \textsuperscript{.}$p<0.1$} 
  %\textsuperscript{*}$p<0.1$}
\end{tabular}
\caption{Individual coefficients' significance and F-test result for various lags.}

\label{tab:granger_results}
\end{table}

\begin{figure*}
    \centering
    \includegraphics[scale=0.6]{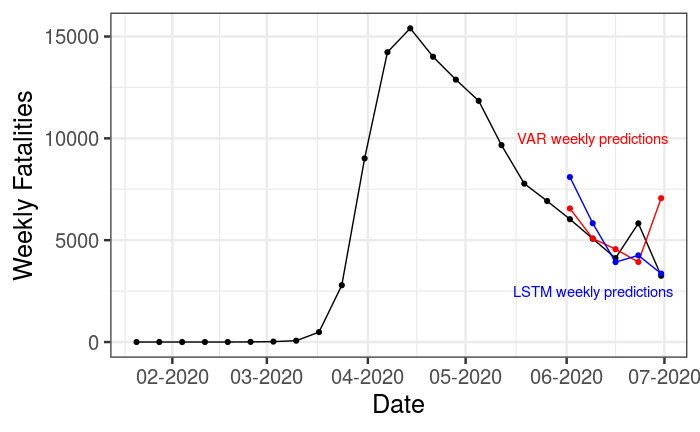}
    \caption{Both models perform reasonably out-of-sample, relative to the amount of data available for learning and the simplicity of the input features.}
    \label{fig:pred}
\end{figure*}

Given the results from our Granger causality analysis we can build a time-series prediction model for estimating the weekly number of fatalities in the near-future (e.g., one week ahead). 
We experiment with two different models, namely, a Vector AutoRegression (VAR) and a Long-Short Term Memory neural network. 
The VAR model is essentially the unrestricted model in the Granger-causality test (Equation \ref{eq:M1}), where $m=p=3$ 
Table \ref{tab:var_model} shows the corresponding model. 
As we can see, increased fraction of people staying home will result in a reduction in the predicted number of fatalities 3 weeks ahead. 
We also examined a stacked LSTM architecture, with 2 layers with 50 hidden units each, followed by a dense layer with ReLU activation. 
We use again a sequence of size of 3 and train the model over multiple epochs using early stopping. 
The results from our two models are presented in Figure \ref{fig:pred}. 
In particular, we provide predictions for the last 5 weeks (as of this writing) and we train each model using all the data up to the week of interest. 
Consequently we make our out-of-sample predictions with each model which are overlaid with the actual fatalities. 
Overall, both models perform relatively well, especially given the short span of the time-series, as well as, the simplicity of the models in terms of input features. 
We would like to note here that these models are not appropriate for longer term predictions (e.g., fatality count in 4 months), which is the focus of most of the fatality-related prediction models developed (\url{https://www.cdc.gov/coronavirus/2019-ncov/covid-data/forecasting-us.html}).

\begin{table}[htbp]\centering
\begin{tabular}{c c c }
\toprule
\textbf{Variable} & \textbf{Coefficient} & \textbf{p-val}\\ 
\midrule
 $\stayHome_{US}(t-1)$         &     $137.9$   &  $0.21$\\
%            &      (65.2)     & \\
 $\stayHome_{US}(t-2)$    &  $252.4$ & $0.15$ \\
%            &   (70.25)        & \\
  $\mathbf{\stayHome_{US}(t-3)}$    &  $-384.7$    & $<0.01$  \\
%            &    (0.21) & \\
{\bf y(t-1) }& $1.35$ & $<0.01$\\
%            &   (0.19)  & \\
y(t-2) & $-0.59$ & $0.16$\\
%            &   (0.19)  & \\
y(t-3) & $0.19$  &  $0.40$\\
%        & 333.1 & \\
\midrule
 $R^2 $          &   0.86  &\\
$SE_{res}$           &  1250  &  \\          
\bottomrule
\addlinespace[1ex]
\multicolumn{3}{l}{\textsuperscript{***}$p<0.01$, 
  \textsuperscript{**}$p<0.05$, 
  \textsuperscript{*}$p<0.1$}
\end{tabular}
\caption{VAR model for predicting weekly fatalities one-week-out.}
\label{tab:var_model}
\end{table}

\section{Quantifying social distancing behavior per demographics}
\label{sec:beta-did-demo}

In the previous section, we saw that there is strong evidence that limiting mobility Granger-causes fewer fatalities from COVID-19. 
Therefore, it is important to understand if and which parts of the population are not able to {\em adhere} to the guidelines. 
This information is critical to be communicated to health officials and policy makers, since it can drive interventions that will help everyone follow the recommendations to the extent possible. 
In this section, we present a framework based on a beta regression model from the daily percentage of time spent home and the difference-in-differences method that can identify the relationship between demographics of interest and the way they relate to social distancing behavior. 

\subsection{Beta regression model}
\label{sec:beta}

Our goal is to model the percentage of time $\stayHome_{\population}$ that a specific population $\population$ spends home daily. 
Given that our dependent variable $\stayHome_{\population}$ is real-valued, bounded in the unit interval a linear regression is not an appropriate model. 
Hence, we choose to use a beta regression model \cite{ferrari2004beta}, where essentially the data are assumed to follow a beta distribution. 
A useful parametrization of the beta distribution for this type of models is given by:

\begin{equation}
    f(y|\mu, \phi) = \dfrac{\Gamma(\phi)}{\Gamma(\mu\phi)\cdot \Gamma((1-\mu)(\phi))} y^{\mu\phi-1}(1-y)^{(1-\mu)\phi - 1}
    \label{eq:beta}
\end{equation}
where $\mu$ is the mean of the beta distribution and $\phi$ is a parameter called precision. 
$\phi$ {\em controls} the variance of the distribution; for a fixed $\mu$, higher precision leads to smaller variance. 
With this setting the beta regression model for $\stayHome_{\population}$ is: 

\begin{equation}
    g(\overline{\stayHome}_{\population}) = \mathbf{x}_{\population}^T\cdot\mathbf{b} + \epsilon 
    \label{eq:beta_reg}
\end{equation}
where $\overline{\stayHome}_{\population}$ is the average daily fraction of time spent home for $\population$, $\mathbf{x}_{\population}$ is the vector of the model's covariates, $\mathbf{b}$ is the vector of the regression's coefficients and $g(\cdot)$ is a link function (strictly increasing and twice differentiable). 
This model is very similar to a generalized linear model (e.g., logistic,  Poisson or negative binomial regression) and it is solved through a Maximum Likelihood Estimation (MLE). 
The MLE identifies the coefficients $\mathbf{b}$, but also the precision parameter $\phi$, which is a constant and not a function of $\mathbf{x}_{\population}$\footnote{There are extensions of this model \cite{simas2010improved} that models the precision as a function of a set of regressions $\mathbf{z}$, i.e., $g^{'}(\phi) = \mathbf{z}^T\cdot\mathbf{c} + \epsilon$.}.

\subsection{Demographics Analysis}
\label{sec:demo}

\begin{figure*}
    \centering
    \includegraphics[scale=0.6]{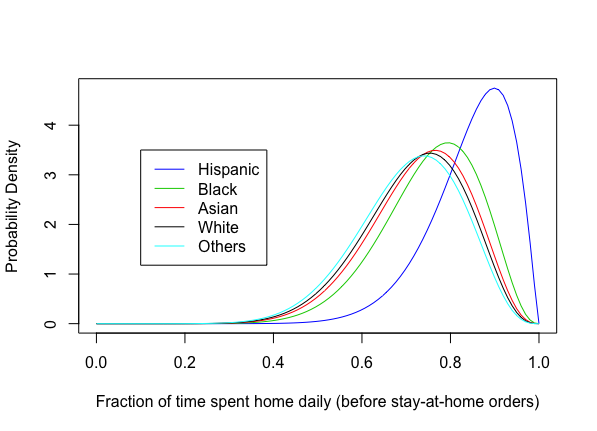}
    \includegraphics[scale=0.6]{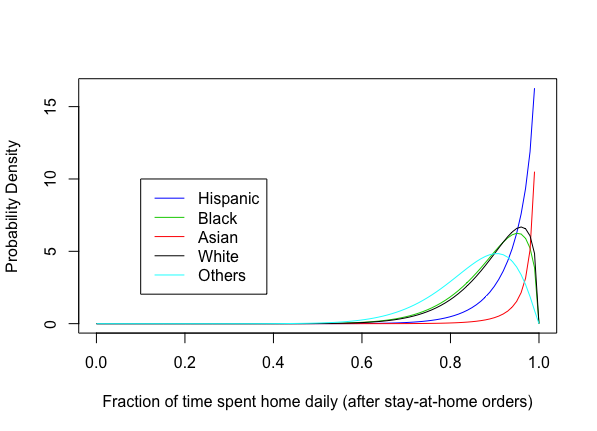}
    \caption{Beta distribution for hypothetical - racially homogeneous - census block group before (left) and after (right) stay-at-home orders across the US. }
    \label{fig:beta-race}
\end{figure*}

In this section we will begin by modeling the fraction of time spent at home daily in each census block as a function of specific demographics of the population. 
We start with race, where census data provide information on the percentage of people within each census block that belong to the following categories: White, Black, Hispanic, Asian, American Indian or Native Alaskan, and Other races\footnote{For the purposes of our analysis we merge the American Indian and Native Alaskan category with the Other races category.}. 
Since we want to estimate the relationship between these demographics and the changes observed after the social distancing recommendations, we build two separate models; one that captures the mobility prior to stay-at-home orders ($M_{pre}$) and one that captures mobility after these orders were put in place ($M_{post}$). 
One of the problems is that different parts of the country put these measures in place in different times through the course of the pandemic. 
Given that the majority of the orders were put in place sometime within March 2020, we build $M_{pre}$ using data from February 2020, and $M_{post}$ using data from April 2020. 
Table \ref{tab:beta_models} presents the results of these regressions. 
Using these results we can start examining the average percentage of time spent daily at home by the population of a hypothetic census block group (HCBG) with a specific racial demographic composition. 
For example, Figure \ref{fig:beta-race} presents the beta distribution for racially homogeneous (hypothetical) census block groups. 
As we can see, there are differences across these hypothetical census block groups, both for the same time period, as well as, their shift as the stay-at-home orders were put in place. 
More specifically, Table \ref{tab:hypothetical-race} presents the average stay home percentage for each of the hypothetical blocks.

\begin{table}[htbp]\centering
\begin{tabular}{c c c }
\toprule
\textbf{Variable} & \textbf{$M_{pre}$} & $M_{post}$\\ 
\midrule
 White\%         &     $-0.45^{***}$   &  $-0.48^{***}$\\
%            &      (65.2)     & \\
Black\%    &  $-0.27^{***}$ & $-0.56^{***}$ \\
%            &   (70.25)        & \\
Hispanic\%     &  $0.29^{***}$    & $0.39^{***}$  \\
%            &    (0.21) & \\
Asian\% & $-0.40^{***}$ & $1.87^{***}$\\
%            &   (0.19)  & \\
Natives+Others\% & $-0.51^{***}$ & $-0.93^{***}$\\
%            &   (0.19)  & \\
constant & $1.39^{***}$  &  $2.5^{***}$\\
%        & 333.1 & \\
\midrule
 $\phi$           &   14.5  & 5.8\\
 N           &  201,917  &   201,917\\          
\bottomrule
\addlinespace[1ex]
\multicolumn{3}{l}{\textsuperscript{***}$p<0.01$, 
  \textsuperscript{**}$p<0.05$, 
  \textsuperscript{*}$p<0.1$}
\end{tabular}
\caption{Beta regression model for the average daily percentage of time of stay home at a census block group before (02/20) and after (04/20) stay-at-home orders.}
\label{tab:beta_models}
\end{table}

\begin{table}[htbp]\centering
\begin{tabular}{c c c }
\toprule
\textbf{Hypothetical Block} & \textbf{Pre} & \textbf{Post}\\ 
\midrule
 White         &  71.8\%   &  89.6\%\\
Black    &  75.6\% &  88.6\%  \\
Hispanic     &  84.4\%    & 94.9\%  \\
Asian & 72.9\% & 98.9\% \\
Natives+Others & 70.7\% & 84.8\%\\
\bottomrule
\end{tabular}
\caption{Percentage of time spent home daily for hypothetical racially homogeneous census block groups based on the beta regression models from Table \ref{tab:beta_models}.}
\label{tab:hypothetical-race}
\end{table}

\begin{figure}
\begin{center}
\begin{tikzpicture}[node distance=0.7cm,inner sep=0.5pt,minimum
size=7mm,scale=0.7]
\tikzstyle{every node}=[]
rectangle (16,12);
\scriptsize

\fill [green!10] (0,4) rectangle (5,10.3);
\fill [green!20] (5,4) rectangle (10.8,10.3);
%\fill [green!10] (10.8,4) rectangle (14,10.3);

\node at (2.5,3.5) (beginning) [date]{02/2020}; 
%\node at (5,1.5) (beginning) [date]{09/18/2008}; 
\node at (5,3.5) (beginning) [date]{03/2020}; 
%\node at (11,1.5) (beginning) [date]{09/18/2010}; 
\node at (7.8,3.5) (beginning) [date, rotate=0]{04/2020}; 
%\node at (14,3.5) (beginning) [date, rotate=0]{04/2013}; 

%March 18, 2011

 \node at (2.5,9.8) () [text=green]{\scriptsize \begin{tabular}{c} Before stay-at-home\end{tabular} }; 
 \node at (8,9.8) () [text=green]{\scriptsize \begin{tabular}{c} After stay-at-home\end{tabular} }; 
 %\node at (12.4,9.8) () [text=green]{\scriptsize \begin{tabular}{c} Purchase verification \\ in   both markets\end{tabular} }; 

% \node at (11,10) () [text=green]{\begin{tabular}{c} After the introduction  of \\ the VP badge\end{tabular} }; 

%%% Eras %%%%%%%%%%
% \draw [decorate,decoration={brace, raise=4pt,amplitude=7pt}, green!90 ,yshift=0pt]
% (0,2) -- (1,2) node [green,midway,xshift=0cm, yshift=0.8cm] {Era 1};
% \draw [decorate,decoration={brace, raise=4pt,amplitude=7pt}, green!90 ,yshift=0pt]
% (1,2) -- (2,2) node [green,midway,xshift=0cm, yshift=0.8cm] {Era 2};
% \draw [green!90 ,yshift=0pt]
% (3,2) -- (4,2) node [green,midway,xshift=0cm, yshift=0.8cm] {...};
% %\draw [decorate,decoration={brace, raise=4pt,amplitude=7pt}, green!90 ,yshift=0pt]
% %(8,2) -- (11,2) node [green,midway,xshift=0cm, yshift=0.8cm] {Era 3};
% \draw [decorate,decoration={brace, raise=4pt,amplitude=7pt}, green!90 ,yshift=0pt]
% (13,2) -- (14,2) node [green,midway,xshift=0cm, yshift=0.8cm] { Era 14};
%%%%%%%%%%%%%%

\draw[->,ultra thick, green] (0,4)--(11.7,4) node[right]{Date};
\draw[->,ultra thick, green] (0,4)--(0,10.5) node[above]{$\overline{\stayHome}_{\population}$};
\draw[-,ultra thick, green] (5,3.9)--(5,4.1) node{};
\draw[-,ultra thick, green] (10.8,3.9)--(10.8,4.1) node{};
%\draw[-,ultra thick, green] (14,3.9)--(14,4.1) node{};
% \draw[-,ultra thick, green] (2,1.9)--(2,2.1) node{};
% \draw[-,ultra thick, green] (5,1.9)--(5,2.1) node{};
% \draw[-,ultra thick, green] (8,1.9)--(8,2.1) node{};
% \draw[-,ultra thick, green] (9,1.9)--(9,2.1) node{};
% \draw[-,ultra thick, green] (10,1.9)--(10,2.1) node{};
% \draw[-,ultra thick, green] (12,1.9)--(12,2.1) node{};
% \draw[-,ultra thick, green] (13,1.9)--(13,2.1) node{};
% \draw[-,ultra thick, green] (11,1.9)--(11,2.1) node{};
% \draw[-,ultra thick, green] (14,1.9)--(14,2.1) node{};

%%% Lines
\draw[-,ultra thick, dotted, UK] (0,6)--(5,6) node [UK,midway, yshift=+0.3cm
.] {$\population_1$};
%\draw[-,thick, UK,dotted] (10.8,7.2)--(14,7) node{};
%\draw[-,ultra thick,dotted, UK] (10.8,7.2)--(14,6.6) node{};

\draw[-,ultra thick, US] (0,5)--(5,5) node [US,midway, yshift=-0.3cm] {$\population_2$};

\draw[-,ultra thick, dotted, UK] (5,8.5)--(10.8,8.5) node [UK,midway, yshift=+0.3cm
] {$\population_1$};

\draw[-,ultra thick, US] (5,8)--(10.8,8) node [US,midway, yshift=-0.3cm] {$\population_2$};
%\draw[-, thick,dotted, US] (5,6.75)--(10.8,6) node{};
%\draw[-,ultra thick, US] (5,6.75)--(10.8,4.7) node{};

%%%

\draw [decorate,decoration={brace, mirror, raise=4pt,amplitude=7pt}, green!90 ,yshift=0pt]
(10.8,8) -- (10.8,8.5) node [green,midway,xshift=0.55cm] {\scriptsize \begin{tabular}{c} $\Delta_2$ \end{tabular}};

\draw [decorate,decoration={brace, mirror, raise=4pt,amplitude=7pt}, green!90 ,yshift=0pt]
(5,5) -- (5,6) node [green,midway,xshift=0.55cm] {\scriptsize \begin{tabular}{c} $\Delta_1$ \end{tabular}};
%\draw [decorate,decoration={brace, mirror, raise=4pt,amplitude=3pt}, green!90 ,yshift=0pt]
%(14,6.6) -- (14,7) node [UK,midway,xshift=0.9cm] {\scriptsize %\begin{tabular}{c} {VP} {effect} \\ {(U.K.)}\end{tabular}};
%\draw [dashed,thick, green!90 ,yshift=0pt]
%(2,7) -- (2.07,8) node [green,midway,xshift=3cm, yshift=-0.25cm, rotate = -4.76] %{\scriptsize \begin{tabular}{c} Parallel  trend assumption\end{tabular}};
%\draw [dashed,thick, green!90 ,yshift=0pt]
%(7.8,6.5) -- (7.87,7.5) node {};
 %\draw [dashed,thick, vpColor ,yshift=0pt]
 %(11.4,6) -- (11.4,7) node [green,midway,xshift=-3cm, yshift=+0.25cm, rotate = -4.76] {};
%%% VP introduction
\draw[<-,ultra thick, vpColor] (5,6)--(5,6.7) node  [vpColor,midway,xshift=0cm, yshift=0.7cm] {\scriptsize \begin{tabular}{c} Stay-at-home \\ orders \end{tabular}};
%\draw[->,ultra thick, vpColor] (10.8,8)--(10.8,7.3) node  [vpColor,midway,xshift=0cm, yshift=1cm] {\scriptsize \begin{tabular}{c} Introduction of  \\ the VP badge \\ on Amazon.co.uk (U.K.)\end{tabular}};
\end{tikzpicture}
\caption{When comparing the mobility post-lockdown for different populations, we need to consider the pre-lockdown mobility as well. }
\label{fig:did}
\end{center}
\end{figure}
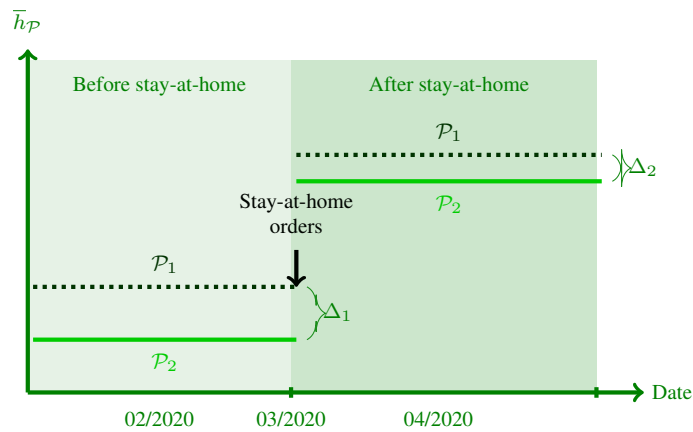

Table \ref{tab:hypothetical-race}, while providing us with a quick view of how specific parts of the population might comply with the social distancing recommendations (in terms of staying home), it does not provide the whole picture. 
In particular, we can see that different demographics are {\em associated} with different levels of mobility outside of their home even before the stay-at-home order. 
So any change observed after the orders were put in place, they need to be compared with the original difference. 
This process is visualized in Figure \ref{fig:did}, where we see two populations $\population_1$ and $\population_2$, with their pre-lockdown daily percentage of staying home, as well as, their post-lockdown daily percentage of staying home. 
While $\Delta_2$ provides us with information about what is happening in the two populations after the stay-at-home orders were put in place, it does not adjust for the {\em behavior} of the two populations prior to the intervention, and the difference $\dd({\population_1,\population_2)}=\Delta_2-\Delta_1$ is more informative. 
Hence, in order to identify demographic discrepancies between two populations, $\population_1$ and $\population_2$, in complying with stay-at-home orders, we performed the following hypothesis test:

\begin{eqnarray}
H_0: & \dd({\population_1,\population_2)} = 0 \label{eq:h0}\\
H_1: & \dd({\population_1,\population_2)} \neq 0 \label{eq:h1}
\end{eqnarray}

In order to perform this test, we use the full beta distribution for each population-time combination and repeatedly sample them to build the distribution of $\dd({\population_1,\population_2)}$. 
Then we can perform the above hypothesis test. 
Table \ref{tab:pval-resample} presents the results for the various comparisons between the minority HCBG and the white one.  
As we can see all minority HCBG - except the Asian one - exhibit a smaller increase as what was expected based on their pre-intervention patterns. 
Particularly interesting is the case of the Hispanic HCBG, which even though exhibits the second highest daily percentage of staying home after the stay-at-home orders, the observed increase is smaller as compared to the white HCBG. 
Furthermore, it is interesting that the Asian HCBG exhibits a 7.5\% higher {\em compliance} as compared to the white HCBG. 
While the reasons for this are not clear - and we cannot identify them through the data we have - there are a few reasons that are plausible, including the increase of racist attacks targeting Asians in the US at the wake of the pandemic \cite{pbs,npr,NYTimes,washpost,usatoday,nbcnews}. 

\begin{table}[htbp]\centering
\begin{tabular}{c c c }
\toprule
\textbf{$\population_1$} & \textbf{$\population_2$} & $\dd(\population_1,\population_2$)\\ 
\midrule
 Black         &     White   &  $-4.8\%^{***}$\\
Hispanic    &  White & $-6.2\%^{***}$ \\
Asian    &  White    & $7.5\%^{***}$  \\
Natives+Others & White & $-3.6\%^{***}$\\
\bottomrule
\addlinespace[1ex]
\multicolumn{3}{l}{\textsuperscript{***}$p<0.01$, 
  \textsuperscript{**}$p<0.05$, 
  \textsuperscript{*}$p<0.1$}
\end{tabular}
\caption{Minority HCBGs exhibit lower percentage of stay-at-home, as compared to white HCBGs.}
\label{tab:pval-resample}
\end{table}

While for the Asian population, staying at home more might also be a way of avoiding racist attacks, the question remains, why are there discrepancies for the rest of the minorities as compared to the white HCBG ? 
One plausible explanation is that a large fraction of these minorities are essential workers and while overall they increase their stay at home, they really need to go to their work. 
Another possible reason is that minorities live in inner cities and as such they are close to their families. 
Furthermore, these minorities have come to rely and support their extended families \cite{taylor2013racial} and hence, they might be providing them with help (e.g., childcare support for essential workers etc.) during this time, leading to higher mobility outside the home. 
Other plausible reasons include the relationship between these groups and technology. 
In particular, ethnic minorities have traditionally been slower in adopting new technology for a variety of reasons \cite{mossberger2012unraveling} and this could mean in a situation like the current pandemic, their inability or unwillingness to use online platforms for essential errands such as grocery shopping. 
While we cannot show with our current data whether any of these plausible reasons are in play, we can examine one additional factor that is relevant to all of the above possibilities; their median income. 
Low income families typically live in inner-city and are of ethnic minorities, they have issues with accessing and adopting technology, while many of the essential workers are low-paid employees (e.g., grocery store workers, delivery, etc.). 
Tables \ref{tab:beta_models-income} and \ref{tab:pval-resample-income} present the same results when we adjust for the median income of an HCBG. 
As we can see, the mobility differences between black and white HCBGs, as well as native and other minorities and white HCBGs, disappears, while for Hispanic and Asian HCBGs the differences are reduced.

\begin{table}[htbp]\centering
\begin{tabular}{c c c }
\toprule
\textbf{Variable} & \textbf{$M_{pre}$} & $M_{post}$\\ 
\midrule
 White\%         &     $-0.43^{***}$   &  $-0.61^{***}$\\
%            &      (65.2)     & \\
Black\%    &  $-0.29^{***}$ & $-0.3^{***}$ \\
%            &   (70.25)        & \\
Hispanic\%     &  $0.27^{***}$    & $0.7^{***}$  \\
%            &    (0.21) & \\
Asian\% & $-0.29^{***}$ & $0.87^{***}$\\
%            &   (0.19)  & \\
Natives+Others\% & $-0.52^{***}$ & $-0.79^{***}$\\
%            &   (0.19)  & \\
Median Income & $-9.9\cdot 10^{-7}$ $^{***}$ & $9.9\cdot 10^{-6}$ $^{***}$\\
constant & $1.43^{***}$  &  $2.13^{***}$\\
%        & 333.1 & \\
\midrule
 $\phi$           &   14.6  & 6.34 \\
 N           &  201,917  &   201,917\\          
\bottomrule
\addlinespace[1ex]
\multicolumn{3}{l}{\textsuperscript{***}$p<0.01$, 
  \textsuperscript{**}$p<0.05$, 
  \textsuperscript{*}$p<0.1$}
\end{tabular}
\caption{Beta regression model for the average daily percentage of time of stay home at a census block group before (02/20) and after (04/20) stay-at-home orders adjusting for median income (expressed in thousands of \$s) in the CBGs.}
\label{tab:beta_models-income}
\end{table}
 
\begin{table}[htbp]\centering
\begin{tabular}{c c c }
\toprule
\textbf{$\population_1$} & \textbf{$\population_2$} & $\dd(\population_1,\population_2$)\\ 
\midrule
 Black         &     White   &  $6\cdot10^{-4}\%$\\
Hispanic    &  White & $-4.3\%^{***}$ \\
Asian    &  White    & $5.7\%^{***}$  \\
Natives+Others & White & $-5\cdot10^{-3}\%$\\
\bottomrule
\addlinespace[1ex]
\multicolumn{3}{l}{\textsuperscript{***}$p<0.01$, 
  \textsuperscript{**}$p<0.05$, 
  \textsuperscript{*}$p<0.1$}
\end{tabular}
\caption{When adjusting for income a large percentage of the mobility differences between HCBGs during stay-at-home orders is explained.}
\label{tab:pval-resample-income}
\end{table}
 
\begin{figure}
 \begin{center}
 \includegraphics[scale = 0.6]{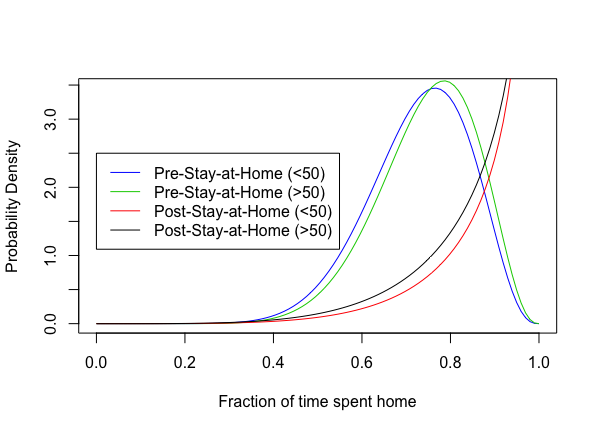}
 \end{center}
 \caption{Beta distributions for the daily fraction of time spent at home for population older and younger than 50.}
 \label{fig:beta-age}
 \end{figure}

We also examined another demographic attribute, namely, age. 
While census provides a breakdown of the age of a census block group in several age brackets, we aggregated them into two bins; younger or older than 50 year old\footnote{Obviously one can repeat the analysis with more bins, but we wanted to keep things simpler mainly for presentation purposes.}. 
Again, we build a beta regression model with the same dependent variable as before but the independent variable is the percentage of the population in the CBG that is older than 50 years old. 
The results are presented in Table \ref{tab:beta_models-age}, where as we can see the older population is associated with a reduced stay-at-home daily time as compared to younger population (less than 50). 
Figure \ref{fig:beta-age} further visualizes the beta distributions for hypothetical CBGs with only population older or younger than 50 years old. 
Furthermore, by performing a similar hypothesis test as in Eq. (\ref{eq:h0})-(\ref{eq:h1}), we find that the HCBG with population older than 50 years old stays at home 2.6\% (p-val < 0.01) \textbf{less} time at home on average as compare to younger population and based on their pre-intervention patterns.  
In contrast to the race case, when adjusting for the median income, the difference remains (-2.5\%, p-val < 0.01). 
A potential reason for this difference between population in the opposite side of the 50 years old mark, can be their {\em technology fluency}. 
Younger people that are avid users of (mobile) technology can take advantage of various services that can help people complete their errands (e.g., grocery shopping), while staying at home. 
This might not be the case for older people (at least to the same extent). 
Again, while this is a plausible mechanism that can drive the observed difference, the data in our disposal does not allow us to further examine this. 

\begin{table}[htbp]\centering
\begin{tabular}{c c c }
\toprule
\textbf{Variable} & \textbf{$M_{pre}$} & $M_{post}$\\ 
\midrule
Older50\% & $0.09^{***}$ & $-0.28^{***}$\\
constant & $0.98^{***}$  &  $2.39^{***}$\\
%        & 333.1 & \\
\midrule
 $\phi$           &   14.2  & 5.5 \\
 N           &  201,917  &   201,917\\          
\bottomrule
\addlinespace[1ex]
\multicolumn{3}{l}{\textsuperscript{***}$p<0.01$, 
  \textsuperscript{**}$p<0.05$, 
  \textsuperscript{*}$p<0.1$}
\end{tabular}
\caption{Beta regression model for the average daily percentage of time of stay home at a census block group before (02/20) and after (04/20) stay-at-home orders based on the percentage of the population that are older than 50 years old.}
\label{tab:beta_models-age}
\end{table}

%%%%%%% tikz image 

% tikzpic.tex
%\documentclass[crop,tikz]{standalone}% 'crop' is the default for v1.0, before it was 'preview'
%\usetikzlibrary{...}% tikz package already loaded by 'tikz' option

\begin{figure*}
    \centering
    \includegraphics[scale=0.4]{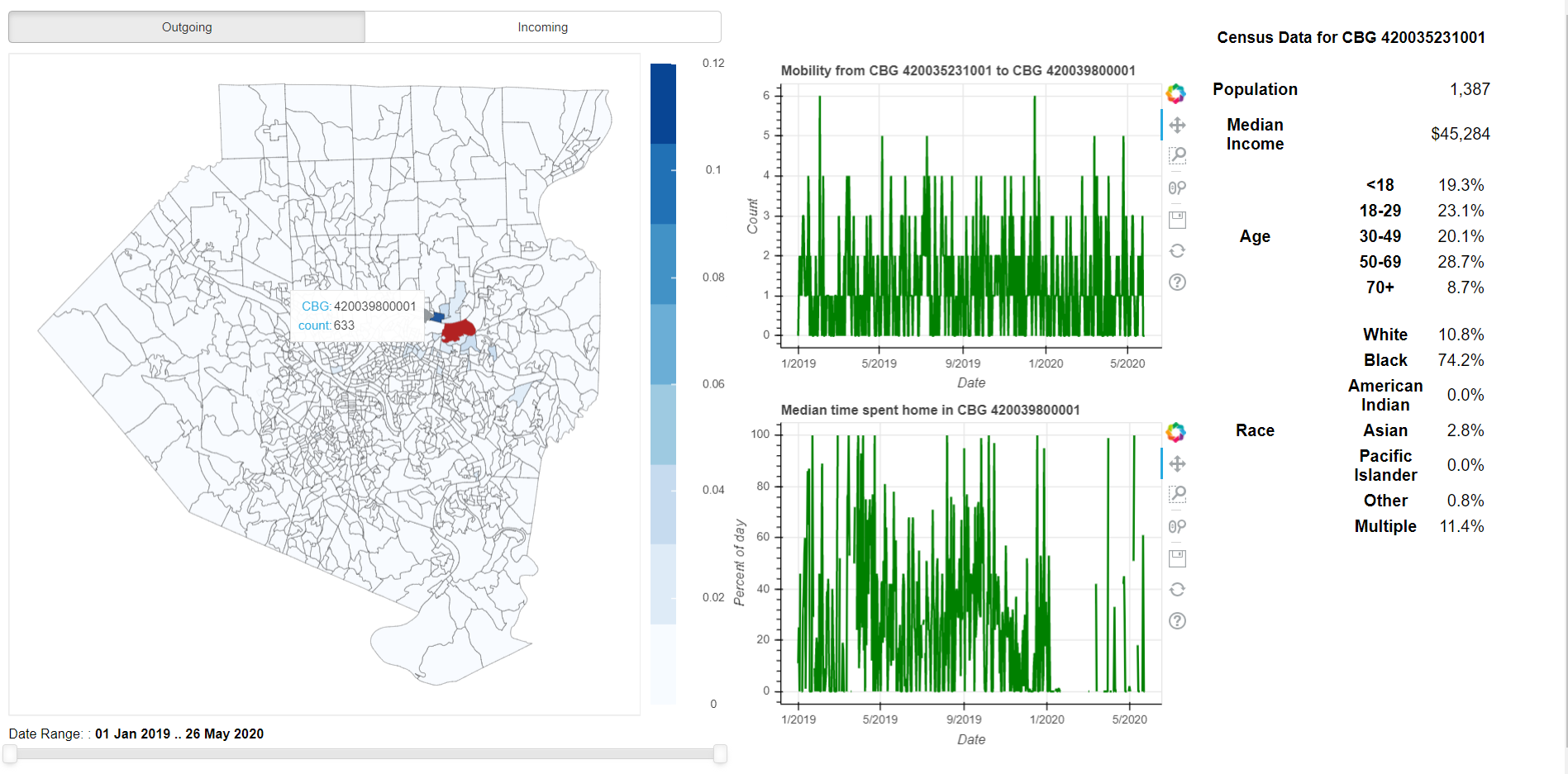}
    \caption{Our dashboard for Allegheny County showing the outgoing mobility from the selected CBG (red) alongside with demographic information. The two time-series plots further provide information related to the interaction between the selected CBG and another CBG that the user hovers over. }
    \label{fig:dashboard}
\end{figure*}

\subsection{Dashboard}
\label{sec:dashboard}

We have also created a dashboard to visualize this mobility information in an interactive  manner\footnote{We will provide a URL with the live dashboard soon.}. 
Figure \ref{fig:dashboard} presents a screenshot from the dashboard that depicts the census block tracts of Allegheny County on the left half. 
The user can choose a tract (the selected tract will be colored red as in the figure) and information about the outgoing mobility (i.e., movements of people whose home CBG is the selected one) and incoming mobility (i.e., movements from people whose home CBG is not the selected one but they visited it) associated with it is presented. 
The choice between outgoing and incoming mobility can be made through the control buttons above the map. 
For example, in Figure \ref{fig:dashboard} outgoing mobility information for people whose home CBG is the selected origin CBG (420035231001) is presented on the map. 
The color for each census block group tract $i$ represents the fraction of the total foot traffic from the residents of the origin CBG, over the period selected from the user\footnote{The user can select the time period through the slider under the map. }, that visited CBG $i$. 
On the right half of the figure, there are two time-series depicted that provide temporal information for the CBG that the user is currently hovering over (say CBG$_h$). 
In the specific situation depicted here, this is CBG 420035231001. 
The top time series provides the daily number of visits in CBG$_h$ from the origin CBG, while the bottom time series represents the fraction of time residents of CBG$_h$ spent at home. 
It is interesting here to note that if we hover over the origin CGB, i.e., CBG$_h$ is the selected CBG, then the top time-series represents {\em self-loops}. 
That is, traffic from residents of the CBG that was destined to other venues/points of interest within the CBG. 
Finally, we also present a table with some basic demographic information about the origin CBG related to our analysis, such as racial and age composition of the population, median income and total population. 

 We would also like to note here that this dashboard is still {\em work-in-progress} in the sense that new features are being added prior to going publicly live. 
For example, our immediate future plan is to visualize information about specific businesses and their geographical reach (i.e., where do customers of different establishments come from?). 
This information can be very helpful for local health authorities when identifying a plan for interventions and the corresponding protocols. 

\section{Conclusions and Discussion}
\label{sec:conclusions}

In this study we perform a macroscopic analysis of the effectiveness of social distancing measures in the US during the COVID-19 pandemic using the notion of Granger causality. 
Our analysis indicate that the average daily fraction of population staying completely at home Granger-causes the number of COVID-19 fatalities in a 3-week period. 
We further examine the presence of bidirectional Granger causality and we do not find any supporting evidence. 
Using this observation, we also build two simple prediction models for weekly COVID-19-related fatalities, using auto-regressive and mobility features.  
We further provide a framework to identify the relationship between demographics and social distancing behavior. 
While this analysis does not provide causal relationships, it can certainly provide important information for policy makers while thinking of ways to increase {\em compliance}. 
Finally, we provide a visualization dashboard with the raw data as well as, the results from our analysis. 
This dashboard is constantly being updated with new results and data. 

We would like to emphasize here that even though we have included a prediction model in our analysis, this is only to showcase in practise the conclusions from the Granger causality analysis\footnote{Furthermore, there are several well-performing prediction models in the public sphere - tracked by CDC as well - and our goal is certainly not to add yet another model.}. 
Furthermore, while the model performs well out-of-sample, several improvements can be achieved by including even more informative features. 
For instance, just an aggregate number of how many hours a person spends out of their home does not capture factors important for the prediction of infections. 
Was this movement to a high-risk location (e.g., a grocery store) or was it for a stroll around the neighborhood? 
Disentangling this is certainly not trivial and we are working in methods for identifying the number of potential  contacts a person from a specific CBG is expected to have based on their mobility and the POI foot traffic data. 
Furthermore, it will be particularly useful to extend our analysis to a more (spatially) fine granularity, focusing on a microscopic analysis (e.g., at the county, or city, level). 
This will allow us to identify the exact time points of interventions and possibly attempt to extract causal relationships using quasi-experimental methods, such as instrumental variables and difference-in-differences.

\bibliographystyle{siamplain}
\bibliography{references}

\end{document}